# Interoperable Framework to Enhance Citizen Services in the Kingdom of Bahrain

Mohammed E. Ghanem
*Information Systems*
*University of Bahrain*
Kingdom of Bahrain
mghanem@uob.edu.bh

Ali Alsoufi
*Information Systems*
*University of Bahrain*
Kingdom of Bahrain
ali.alsoufi@gmail.com

*Abstract* - Citizen records are scattered between different state organizations. It wastes time, effort, and resources for both citizen and organization to collect, maintain, and update records to fulfill citizen services. Interoperability is a key element that enables seamless collaboration between different entities. It requires non-conventional methods to overcome interoperability challenges such as lack of trust, centralization, and policy and technology differences. Blockchain is a disruptive technology with the potential to overcome these challenges. The technology designed to enable peer-to-peer transactions with elimination of intermediary in a trustless environment through the control of consensus mechanisms. This research aims to explore the status of interoperability in Bahrain, design an interoperable framework, and then test the validity of the framework by implementation of a prototype using blockchain technology. The research will be divided into four phases; I: Information collection, II: Design and modeling the framework, III: Implementation of a prototype, and Phase IV: Measuring the performance of the prototype. This research is in progress and it is expected, once is it complete, to enhance the e-government's plan in the Kingdom of Bahrain to provide better services to citizens and help in the transition from e-government to seamless government, which will lead to sustainable citizen services. On the other hand, the findings of the study is expected to improve the social, economical, and environmental sustainability by the increase in process optimization, reduction of cost and complexity.

## I. INTRODUCTION

Bahraini e-government's main purpose is to enable citizens to conduct public and private transactions in a secure and reliable means using minimum resources. However, the reality is the services currently provided by the e-government is adequately effective in terms of maintaining time, effort, and money, but within the scope of the competent authority to complete only. The key question to ask is "what about services require official documents from several sources before being used by the competent authorities". For example, if the citizen wants to submit a request to the Ministry of Housing, he/she must [1]:

1. Present a valid identity card copies for both husband and wife that includes the employer, so if the current employer is not available the citizen must first acquire physically an official employment letter from the employer then visit the Central Informatics Organization in person to update his/her identity card.
2. An official certification of currently owned properties, so again the citizen will waste time, effort and money to visit ministry of housing in person to acquire that certification.
3. A Benefit report, which will require to book an appointment and visit the (Benefit company) to receive one.
4. An official one-month valid income letter (more paper documents).
5. Copies of birth certificates.
6. Copies of passports.

The previous steps show the amount of the records that the citizen needs to collect and the various organizations that he/she has to deal with, thus highlight number of problems such as loss of documents or error in completing them. Moreover, after the long-suffering of the effort and time and resources incurred by the citizen and the concerned authorities and after the submission of the request successfully other problems start to emerge, such as updating the data, following up on the request, which introduces the citizen into an endless circle of compilation and updating of records from various parties.

The citizen register is one of the most important records that the government maintains, monitors and updates, because of its importance to the conduct of majority of public and private services transactions every day, beginning with the birth certificate and through the identification card and the passport ending by the death certificate [2]. Each government entity has its own type of records either short-term (renewable), and long-term (e.g. birth certificate) and these records need a secure immutable sustainable storage system, as well as the own organization standards to preserve and update. However, citizen records are scattered and inconsistent throughout different state organizations. Lots of paper wastage and difficulty to information access. Valuable time and efforts wasted in maintaining and updating citizen's records. On the other hand, citizens struggle to own their public records and keep it up to date. Many countries suffer the same problems because of barriers such as trust among different organizations, centralization, and policy and technology incompatibility [3].

A highly matured e-government is an interoperable one. There is an obvious lack of interoperability in e-government, and this lack exists due to the difficulty of determining a central authority to information. To share information is not only a matter of equipping the right technology; it is a combination of policy, management, and technological capabilities [4]. As Pardo, Nam, & Burke [4] concluded, non-conventional methods like decentralization could lead the way to successful implementation of e-government interoperability (eGI).

Centralization, lack of trust, records tampering, and lack of transparency are the major challenges that blockchain technology can solve. In 2008, the anonymous Satoshi Nakamoto initiated a decentralized foundational technology called blockchain. Later in 2009, the originally first cryptocurrency (Bitcoin) was released to originate a revolutionary non-conventional technology, this technology designed as a response to the problems led to the financial crises 2007-2008 [5].

Blockchain is series of connected hashed blocks, each block contains but not limited to transactions data, the block hash key and the previous block's (parent block) hash key, and timestamp. The first block on the chain is called genesis block. The following example illustrates the creation and approval of blocks; first user (A) initiates a transaction by sending a certain amount of cryptocurrency to user (B), once the transaction initiated it become public to all the network users. Second, certain network nodes called (miners) will authenticate the transaction by confirming the sender's data through solving a mathematical puzzle, and this process is called (mining). The transaction will be added to a block once authenticated, then when the block capacity of authenticated transactions reached the block is timestamped and added to the chain [6]. The assembled chain is similar to a public distributed ledger where every node on the network holds an updated copy. Blockchain technology roughly implemented within three structures; 1- public blockchain network, where any user can join and any node can be part of the mining process, 2- consortium blockchain network, joining is almost public however the nodes must be permissioned to join the mining process, 3- private blockchain network, limited membership and mining is done through one controlling node (Zheng, Xie, Dai, Chen, & Wang, 2018).

Until recently, the focus was on Bitcoin and cryptocurrency research, but, with the advancement in the technology, there has been an increase in researches exploring the use of blockchain in other fields such as healthcare, energy, public and government services, and supply chain management [7]. All previous fields' demands collaboration between different parties to reach optimum performance.

In healthcare, the integration of blockchain technology will enable different consultation teams' within different entities to share their opinions about cancer cases, which will provide the most suitable treatment for the patient [8]. Blockchain and smart contracts can play great role in enhancing the process of the Local energy markets (LEM), which is a project proposed by the European Commission in 2016 by utilizing the immutability and the transparency of blockchain to record real time meter data and billing info, which will increase micro and macro grid efficiency, and reduces energy waste [9], [10].

## II. RESEARCH OBJECTIVES

The development of public services for citizens in terms of ease of access and speed of the process has become an urgent matter for e-governments so they can transform from service providing to citizen needs fulfillment [11]. E-governments meant to improve citizen's life by delivering better high quality services, increase citizens' satisfaction, enhance time and cost efficiency [12]. E-government Interoperability is one of the main capabilities to help improve citizen's public services [3]. In order to achieve interoperability capability conventional methods are not the answer, as they are very complex and costly to apply [4], [13]. Pardo, Nam, & Burke [4] and Paul & Paul [13], argued that some of the barriers to implement eGI using conventional methods such as; difficulties to change organizations infrastructure, change in organizational policies, and employees are usually resistant to the adoption of new work environments (procedures and technologies).

In order to improve the interoperability of the Bahraini e-government the following research objects must be achieved:

- To evaluate the status of interoperability between state organizations in Bahrain.
- To develop a framework to a unified portable citizen's public and private records using blockchain technology.
- To evaluate the Blockchain technology capabilities in simplifying the public and private services to the citizens in Bahrain.
- To develop a prototype, as a proof of concept for the suggested framework.

## III. METHODOLOGY & FRAMEWORK

The government and private institutions represent the ecological and vital system of any country and cooperation between them provides the infrastructure to better serve the citizens. The study will be conducted on one of the public e-services provided by the e-government portal as a case study through the creation of an interoperable network between the organizations involved in the transaction process. This research will be divided into four phases as following:

*A. Phase I: Information collection*

Information collection through conducting semi-structured interviews with involved organizations to better understand the current process and available systems. An exploratory research by the researcher to estimate the usual time and cost required per citizen to complete a transaction's full cycle.

*B. Phase II: Design and modeling the framework*

In this phase, a comprehensive framework will be developed to support my research objectives. Based on the initial study conducted, the following high-level conceptual framework (Fig. 1) will be used. It is based and focused on three pillars: 1- security and privacy of the records for both citizens and organizations, 2- reliability and authenticity of the records and the process, 3- the record keeping standards for collecting, processing, storing, and dissemination.

The conceptual framework demonstrates the interaction between different stakeholders of records and how the blockchain network will enhance the collaboration between them. The citizen will be responsible of acquiring and maintaining the eKey service, which is provided by the e-government authority. The eKey must be used to request services securely and to confirm the identity of the requester. The e-government is responsible to generate eKeys to citizens, and to authenticate theses eKeys to initiate requested services. The blockchain will designate the smart contract related to the initiated service to check for documents needed to fulfill the requested service. The citizen must authorize the document collection process. Once the required documents are fulfilled, the smart contract will authorize the target service provider to collect a copy of these documents and the service will be completed. On the other hand, once a service provider issues a new document, a copy of that document will be added to the blockchain network where other organizations must validate it through the mining process. The original document will be stored on the service provider's storing facility to grantee availability and authenticity. All stakeholders shall keep a copy of the ledger, however they are authorized through smart contracts only to preview documents needed to fulfill services they provide. The citizen will have access to his/her related documents anytime and anywhere.

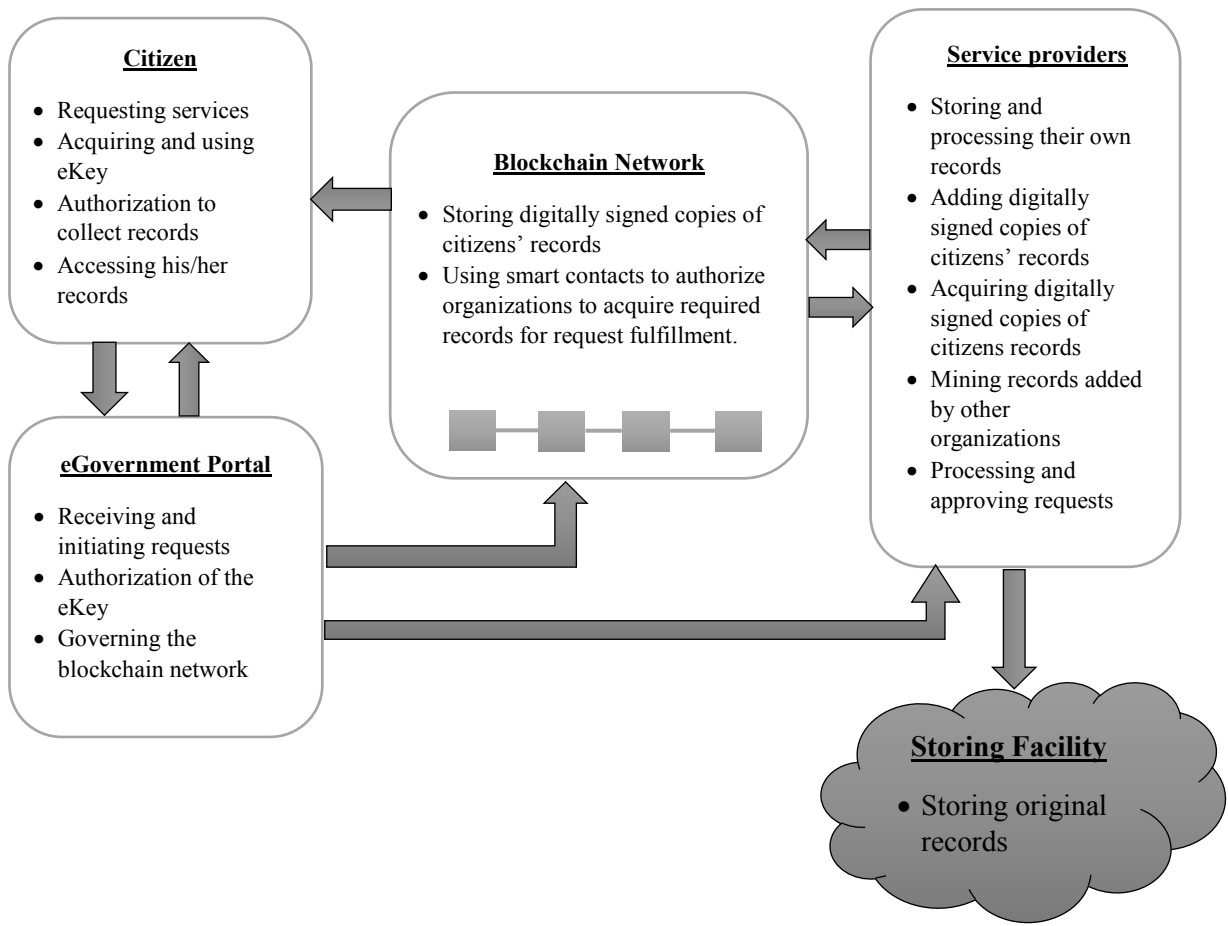

Fig. 1. High-level conceptual framework to identify the relationship between the entities involved in the service process and the blockchain network.

## C. Phase III: Implementation of a prototype

A prototype will be implemented and tested through the phases of the Systems Development Life Cycle (SDLC), the SDLC includes but not limited to analysis, design, implementation, and maintenance stages. The analysis stage depends on the research's first phase, where the current process will be analyzed based on the semi-structured interviews. The system design stage will include the data flow diagrams and entity relationship diagrams to simulate the actual interaction of the system entities. The implementation will be on one of the available blockchain netwroks such as Etherium or through the e-government network, and by using consortium blockchain to control the nodes allowed joining the mining process. The platform will be built and tested through web/mobile interface.

Many models and methodologies of SDLC are available and efficient, however based on the nature of the study, using the evolutionary prototyping model will be more effective.

## D. Phase IV: Measuring the performance of the prototype

Measuring the effect of process speed and cost saving through conducting post implementation qualitative research. The researcher will run a full cycle process through using the prototype to apply for the selected public e-service to estimate the usual time and resources required. The collected data will be analyzed and compared with the pre-implementation data.

## IV. CONCLUSION

Interoperability is a key factor in measuring the maturity of e-government services. Furthermore, interoperability is a tool that will enable seamless services between state organizations where citizens will be able to control and maintain their records. Blockchain is a disruptive technology that still immature, however it demonstrated great potential in fields such as healthcare, energy supply, supply chain, and public and governmental services.

This is a research in progress and the outcome is expected to enhance the e-government's plan in the Kingdom of Bahrain to provide better services to citizens and to help in the transition from e-government to seamless government, which will lead to sustainable citizen services. The proposed framework is expected to have minor impact on the changes in the current e-government policies and framework.

The findings of the study will advance the social, economical, and environmental sustainability through the increase in process optimization, reduction of cost and complexity. On the other hand, this study will help uncover critical research areas related to interoperability in Bahrain.

REFERENCES

[1] "Bahrain. Ministry of Housing," 2017. [Online]. Available: http://www.housing.gov.bh/ar/EServices/ApplyNewApplication/Pages/ApplyForNewUnit.aspx.

[2] "Bahrain. eGovernment," 2018. [Online]. Available: https://www.bahrain.bh.

[3] P. Gottschalk and H. Solli-Sæther, "Interoperability in E-Government: Stages of Growth," in *Integrating E-Business Models for Government Solutions: Citizen-Centric Service Oriented Methodologies and Processes*, 2009, pp. 50-66.

[4] T. A. Pardo, T. Nam and G. B. Burke, "E-Government Interoperability : Interaction of Policy, Management, and Technology Dimensions," *Social Science Computer Review,* 2012.

[5] M. Iansiti and K. R. Lakhani, "The Truth About Blockchain," 2017. [Online]. Available: https://hbr.org/2017/01/the-truth-about-blockchain.

[6] Z. Zheng, S. Xie, H.-N. Dai, X. Chen and H. Wang, "Blockchain challenges and opportunities: A survey," *International Journal of Web and Grid Services,* vol. 14, no. 4, pp. 352-375, 2018.

[7] V. Gatteschi, F. Lamberti, C. Demartini, C. Pranteda and V. Santamaría, "To Blockchain or Not to Blockchain: That Is the Question," *IT Professional,* pp. 62-74, 2018.

[8] P. Zhang, D. C.Schmidt, J. White and G. Lenz, "Blockchain Technology Use Cases in Healthcare," *Advances in Computers,* vol. 111, pp. 1-41, 2018.

[9] C. Zhanga, J. Wua, C. Longa and M. Chenga, "Review of Existing Peer-to-Peer Energy Trading Projects," *Energy Procedia,* vol. 105, pp. 2563-2568, 2017.

[10] E. Mengelkamp, B. Notheisen, C. Beer, D. Dauer and C. Weinhardt, "A blockchain-based smart grid: towards sustainable local energy markets," *Computer Science - Research and Development,* vol. 33, no. 1-2, p. 207–214, 2018.

[11] R. M. Davison, C. Wagner and L. C. Ma, "From government to e- government: a transition model," *INFORMATION TECHNOLOGY & PEOPLE,* pp. 280-299, 2005.

[12] U. A. Mokhtar and Z. M. Yusof, "Records management practice: The issues and models for classification," *International Journal of Information Management,* pp. 1265-1273, 2016.

[13] A. Paul and V. Paul, "A Framework for e-Government Interoperability in Indian Perspective," *International Journal of Computer Information Systems and Industrial Management Applications,* pp. 582-591, 2014.